\documentclass[aps,pra,reprint, amsmath, amssymb,superscriptaddress]{revtex4-1}

\usepackage{bm}
\usepackage[retainorgcmds]{IEEEtrantools}
\usepackage{graphicx}
\usepackage{mathrsfs}
\usepackage{amsmath}
\usepackage{amssymb}
\usepackage{color}
\usepackage{amsfonts}
\usepackage{times,txfonts}
\usepackage{nicefrac}
\usepackage{physics}

\newcommand{\bs}{\bm{\sigma}}
\newcommand{\ud}{\,\mathrm{d}}

\newcommand{\trans}{^{\text{T}}}

\newcommand{\wk}{\omega_k}

\begin{document}

\title{Exact open dynamics of a generalized class of dephasing-type spin-boson models}
\date{\today}
\author{Mariana Afeche Cipolla}
\affiliation{Instituto de F\'isica da Universidade de S\~ao Paulo,  05314-970 S\~ao Paulo, Brazil}
\author{Gabriel T. Landi}
\email{gtlandi@if.usp.br}
\affiliation{Instituto de F\'isica da Universidade de S\~ao Paulo,  05314-970 S\~ao Paulo, Brazil}

\begin{abstract}

We discuss  a generalization of the dephasing-type spin-boson model in which $N$ qubits are connected to $K$ bosonic modes in an arbitrary way. 
The model can be solved exactly for any initial state and coupling type. 
We found that this leads to two important effects, which are specific of the multipartite nature of the problem. 
First, the bosons can mediate an effective interaction among the spins, which corresponds to a type of time-dependent Lamb-shift. 
By tuning the types of couplings, this interaction can also be varied at will. 
Second, the dephasing rate of the qubits is found to depend on the order of the coherence, with superpositions that have drastically different magnetizations being exponentially more affected. 
The relative strength of the Lamb-shift and dephasing can also be tuned by changing the spin-boson couplings.

\end{abstract}
\maketitle{}

%
%
\section{\label{sec:int}Introduction}
%
%

Open quantum systems are ubiquitous in physics and play a central role in the development of new quantum technologies. 
All systems are, to one degree or another, in contact with its surroundings. 
The correlations that develop due to these interactions deteriorate the information-theoretic properties of the system.
Isolating it from this deleterious contact is therefore one of the major challenges in the development of quantum-coherent experimental platforms~\cite{Haroche2006}. 
For this reason, understanding open system dynamics remains a timely and important problem in the field. 

Due to the complex  structure of realistic environments and  system-environment interactions, however, there are very few models of open system dynamics which can be solved exactly~\cite{Breuer2007}. 
This is unfortunate. 
Most of the modeling done on open quantum systems rely on several approximations (Born-Markov, etc.) which are often uncontrolled and may lead to dramatic consequences. 
One example is the ``local~vs.~global'' dilemma~\cite{Correa2019,Gonzalez2017,Levy2014,Mitchison2018,Ribeiro2015,Rivas2010b,Wichterich2007} surrounding Lindblad master equations for systems composed of multiple parts (such as e.g. spin chains): local equations, derived phenomenologically, do not properly thermalize the system and have been argued to violate the second law~\cite{Levy2014,DeChiara2018}. Global master equations, on the other hand, are highly non-local and gives rise to unphysical heat currents~\cite{Wichterich2007,Santos2016}. 
This conundrum and all proposals to correct it, such as e.g. Redfield equations \cite{Purkayastha2016}, is a consequence of the uncontrolled approximations involved in deriving the master equations. 

In this sense, exactly soluble models can offer a fresh new perspective on the problem, as they allow one to have full control over the system-bath dynamics at all time scales. 
Thus, although usually idealized, they provide valuable insight into this difficult problem. 
Among the class of exactly soluble models, one which is particularly famous, is the dephasing-type spin-boson model, where a single qubit interacts with $K$ bosonic modes (characterized by annihilation operators $b_k)$ through the Hamiltonian~\cite{Palma1996}
\begin{equation}\label{SB_basic}
H = \frac{\epsilon}{2} \sigma_z + \sum\limits_{k=1}^K \bigg\{ \omega_k b_k^\dagger b_k + \lambda_k \sigma_z (b_k + b_k^\dagger) \bigg\}, 
\end{equation}
where $\sigma_z$ is the Pauli matrix for the qubit. 
Since the interaction involves only $\sigma_z$, it does not cause any transitions in the computational basis of the qubit. 
It's only effect is therefore to cause decoherence.

This model has several extremely nice features. 
Decoherence means the off-diagonal element will evolve with an additional term $e^{-\Gamma(t)}$, where $\Gamma(t)$ is a function of the qubit-boson coupling, as well as the initial state of the bosons. 
In the case where the bosons start in a thermal state, it reduces to~\cite{Breuer2007}.  
\begin{equation}\label{SB_Gamma}
\Gamma(t) = \sum\limits_{k=1}^K \frac{4 \lambda_k^2}{\omega_k^2} \big[1- \cos(\omega_k t)\big] \coth\left(\frac{\omega_k}{2T}\right). 
\end{equation}
This expression is exact. 
It can be used for the case where the number of bosons $K$ is finite or can be converted to an integral when $K\to \infty$. 
It therefore provides with ways to analytically explore effects such as non-Markovianity (which depends on the value of  $K$) and the typical time-scales of the problem (which depend sensibly on the temperature $T$). 

\begin{figure}[!t]
\centering
\includegraphics[width=0.4\textwidth]{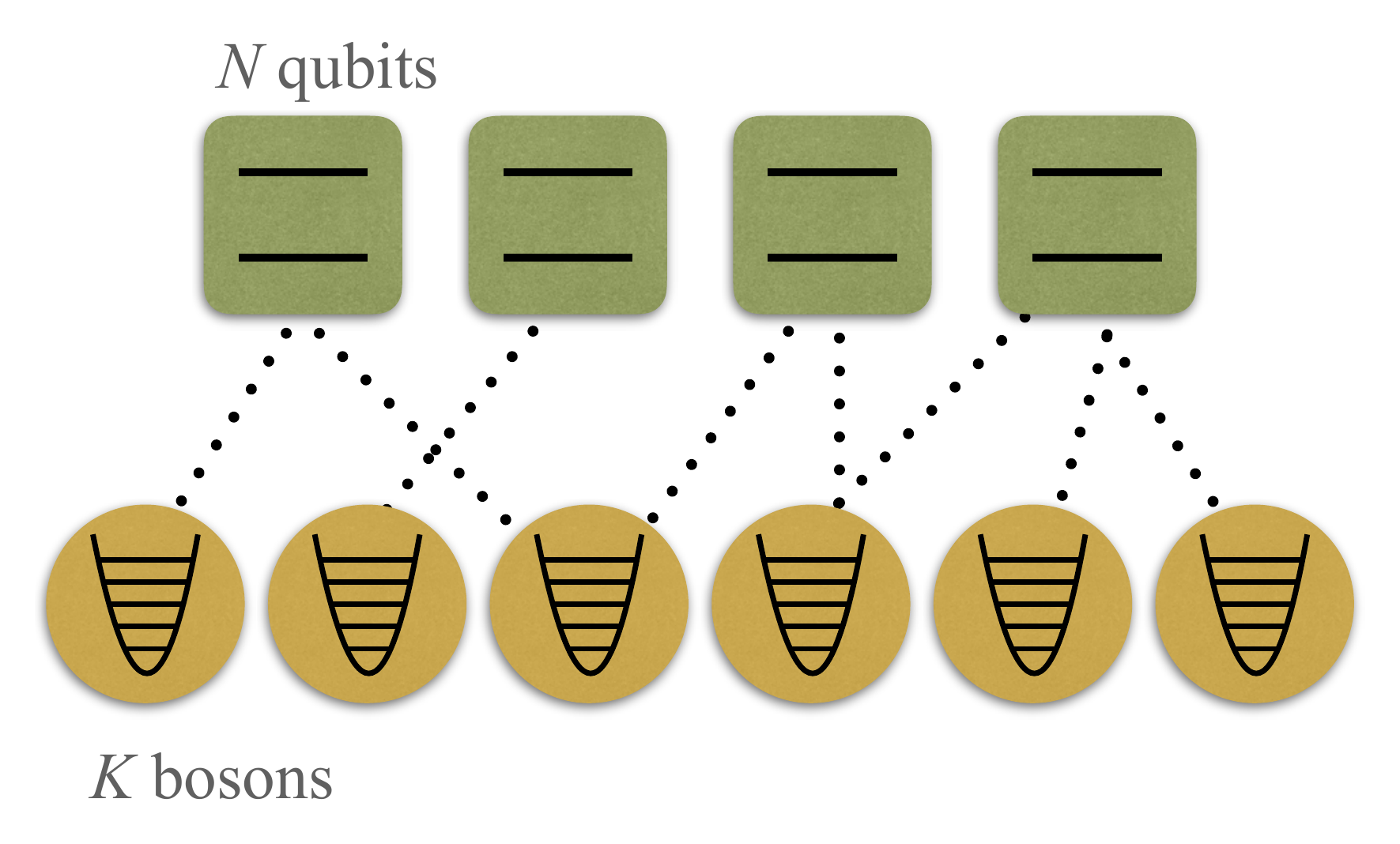}
\caption{\label{fig:diagram}
Generalized spin-boson model where $N$ qubits interact in an arbitrary way with $K$ bosons according to the dephasing-type interaction~(\ref{H}). 
}
\end{figure}

Because it is analytically soluble, the spin-boson model~(\ref{SB_basic}) has found multiple applications, from quantum computing \cite{Palma1996} and non-classicality \cite{Friedenberger2018} to metrology \cite{Walborn2018,Razavian2018} and biomolecular physics \cite{Nalbach2010}.
It is also well within reach of several quantum platforms, such as trapped ions \cite{Wu2013a} or superconducting qubits \cite{Makhlin2001}.
Some variations of this model have also been analyzed,  for instance by assuming that the bosons are prepared instead in a squeezed state~\cite{You2017}. 
The features which enable the model to be analytically solved, however, are much more general. 
There is, therefore, a much broader class of models which can also be solved analytically using similar methods. 

In this paper we exploit this to the limit and discuss the general solution for a broad class of dephasing-type spin-boson models. 
More specifically, we consider a system composed of $N$ qubits (henceforth referred to as $S$) and $K$ bosons (henceforth referred to as $B$), interacting with a Hamiltonian of the form (see Fig.~\ref{fig:diagram}): 
\begin{IEEEeqnarray}{rCl}
\label{H}
H &=& H_S(\bs_z) + V(\bs_z),	\\[0.2cm]
V(\bs_z) &=&  \sum\limits_{k=1}^K  \bigg\{\omega_k b_k^\dagger b_k + f_k(\bs_z) (b_k + b_k^\dagger)\bigg\}. 
\label{V}
\end{IEEEeqnarray}
Here $H_S$ is the Hamiltonian for the qubits, which we do not need to specify exactly. 
All we assume is that $H_S$ is a function only of the Pauli $\sigma_z^i$ operators, $\bs_z = (\sigma_z^1, \ldots, \sigma_z^N)$. 
This includes, for instance, the case of a classical Ising Hamiltonian $\sigma_z^i \sigma_z^j$. 
In addition, the interaction with the Bosons is  assumed to be characterized by a generic function $f_k(\bs_z)$, which is also only a function of the $\sigma_z^i$. 
As a consequence, it follows that $[H_S,V] = 0$, so that the interaction does not cause any changes in the populations of the qubits in the computational basis. 

The role of $V$ will therefore be to only cause decoherence in the computational basis. 
However, when multiple qubits are present, the notions of decoherence become much more subtle. 
Quantum coherences in the multipartite scenario can be classified in different ways.
It can, for instance, be local or global: local coherence is that which is also present in the reduced density matrices of the individual qubits, whereas global coherences exist only in the composite state \cite{Streltsov2016a,Baumgratz2014,Kraft2018,Winter2016,Regula2018}. 
Global coherences can also be further classified using the notion of coherence orders~\cite{Pires2017}. 
As we will show, the model~(\ref{H}) allows one to neatly explore these difference concepts. 

The model~(\ref{H}) can be solved exactly for any function $f_k(\bs_z)$, with an arbitrary number of spins and bosons.  
This structure therefore contemplates a large range of physically interesting scenarios. 
For instance, suppose that the $f_k$ is a linear function 
\begin{equation}\label{f_linear}
f_k(\bs_z) = \sum\limits_{i=1}^N \lambda_{ki} \sigma_z^i,
\end{equation}
for some set of coefficients $\lambda_{ki}$ which measure the coupling strength from spin $i$ to boson $k$. 
This corresponds to a direct generalization of the model~(\ref{SB_basic}). 
The difference is that, now, multiple qubits may interact with the same bosonic mode. 
As a consequence, the bosons can mediate an effective interaction among the qubits. 
The coupling~(\ref{f_linear}), for instance,  leads to an effective interaction of the Ising type $\sigma_z^i \sigma_z^j$, with arbitrarily tunable coupling strengths. 

We can also consider particular cases of the linear interaction~(\ref{f_linear}). 
One such case is when bosons are coupled to the qubits in sectors, such that each qubit only interacts with its own set of bosons. 
This can be built, for instance, by taking $K = m N$, where $m$ is an integer. 
We then set $\lambda_{k1} \neq 0$ for $k = 1,\ldots, m$, $\lambda_{k2} \neq 0$ for $k = m+1, \ldots, 2m$ and so on. 
In this case each qubit will have its own independent dephasing channel, so that there will be no communication among the qubits through the bosons. 

We can also consider the opposite scenario where a single boson ($K = 1$) interacts with equal strength with all qubits. 
The interaction~(\ref{V}) in this case becomes
\begin{equation}\label{Dicke}
V(\bs_z) = \omega b^\dagger b +  \lambda \bigg(\sum\limits_{i=1}^N\sigma_z^i \bigg) (b+ b^\dagger). 
\end{equation}
This is the dephasing-analog of the Dicke model~\cite{Dicke1954}, where $N$ qubits interact collectively with a single bosonic mode. 

One is not restricted to linear interactions such as~(\ref{f_linear}), however.
Eq.~(\ref{V}) can also contemplate 3-body or higher interactions. 
For instance, $f_k(\bs_z)$ can involve arbitrarily long strings of Pauli-Z operators, $\sigma_z^i \sigma_z^j \sigma_z^n \ldots$. 
The model~(\ref{H}) can also be solved exactly for arbitrary initial states of $S$ and $B$. 
This even includes, for instance, situations where the bosons start in a correlated state. 
Below we will discuss this general case. As we show, the solution becomes particularly nice if we assume that the initial state of $B$ is Gaussian and therefore fully characterized by a covariance matrix~\cite{Serafini2017}. 
The complete solution of Eq.~(\ref{H}) is provided in Sec.~\ref{sec:sol}. 
The particular case of linear interactions, Eq.~(\ref{f_linear}), is discussed in Sec.~\ref{sec:cases}. 
Finally, the relevance of our results to present-day applications in quantum information sciences is  discussed in Sec.~\ref{sec:disc}.

%
%
\section{\label{sec:sol}General solution of the multi-qubit spin-boson model}
%
%

We consider here the model described by Eq.~(\ref{H}). 
We work in the interaction picture with respect to $H_S(\bs_z)$. Since $[H_S,V] = 0$, this does not affect $V$. 
The qubits and bosons are initially prepared in arbitrary states $\rho_S$ and $\rho_B$.
After a time $t$, their joint state $\rho_{SB}(t)$ will then be given by 
\begin{equation}\label{global_map}
\rho_{SB}(t) = e^{-i V(\bs_z) t} \rho_S \rho_B e^{i V(\bs_z) t}. 
\end{equation}
We shall adopt here a spin notation for the computational basis of the qubits. 
That is, we write the eigenstates of $\sigma_z^i$ as $|\sigma_i\rangle$ where $\sigma_i = \pm 1$ are the corresponding eigenvalues. 
The usual computational basis is mapped as $|0\rangle = |\sigma = +1\rangle$ and $|1\rangle = |\sigma = - 1\rangle$. 
Moreover, the joint state of the $N$ qubits will be denoted by 
$|\bs \rangle = |\sigma_1, \ldots, \sigma_N\rangle$.

\subsection{General form of the dephasing map}

Taking the trace of Eq.~(\ref{global_map}) with respect to the bosons leads to the map
\begin{equation}\label{local_map}
\rho_S(t) = \mathcal{E}_t(\rho_S) = \tr_B \bigg\{ e^{-i V(\bs_z) t} \rho_S \rho_B e^{i V(\bs_z) t}\bigg\}. 
\end{equation}
We now exploit the fact that $V(\bs_z)$ is already diagonal in the computational basis, so that
\begin{equation}
V(\bs_z) |\bs\rangle = V(\bs) |\bs\rangle, 
\end{equation}
where $V(\bs)$ is identical to the operator $V(\bs_z)$ in Eq.~(\ref{V}),   with $f_k(\bs_z)$  replaced by $f_k(\bs)$:
\begin{equation}\label{V_func}
V(\bs) =  \sum\limits_{k=1}^K  \bigg\{\omega_k b_k^\dagger b_k + f_k(\bs) (b_k + b_k^\dagger)\bigg\}. 
\end{equation}
$V(\bs)$ is therefore a $c$-number with respect to the qubit sector, but still an operator for the bosons.

Using this fact, the matrix elements of the map~(\ref{local_map}) become
\begin{IEEEeqnarray*}{rCl}
\langle \bs |\mathcal{E}_t(\rho_S) | \bs\rangle 
&=& \tr_B \bigg\{e^{- i V(\bs) t} \langle \bs | \rho_S |\bs'\rangle  \rho_B e^{i V(\bs) t}\bigg\}		\\[0.2cm]
&=& \langle \bs | \rho_S |\bs'\rangle \langle  e^{i V(\bs) t}e^{- i V(\bs) t} \rangle_B.
\end{IEEEeqnarray*}
This motivates us to define
\begin{equation}\label{Lambda}
 \Lambda_{t}(\bm{\sigma}, \bm{\sigma}') =\log{\langle e^{i V(\bm{\sigma}') t} e^{-i V(\bm{\sigma}) t}  \rangle_B}, 
\end{equation}
so that the map~(\ref{local_map}) can be written, element-by-element, as 
\begin{equation}\label{local_map_elements}
\langle \bs | \mathcal{E}(\rho_S) | \bs' \rangle = \langle \bs | \rho_S | \bs' \rangle e^{\Lambda_{t}(\bm{\sigma}, \bm{\sigma}')},
\end{equation}
Note, in particular, that $\Lambda_t(\bm{\sigma},\bm{\sigma}) = 0$, so that the populations in the computational basis are not affected by the interaction with the bosons.

Next, using the basic algebra of bosonic coherent states, one may show that 
\begin{equation}
e^{i V(\bs') t} e^{-i V(\bs) t} = e^{i W_ t(\bs')} e^{-i W_t(\bs)} \prod\limits_k  D_k\left(\mu_k(\bs, \bs')\right),
\end{equation}
where $D_k(\alpha) = e^{\alpha b_k^\dagger - \alpha^* b_k}$ is the displacement operator, 
\begin{equation}\label{mu_k}
    \mu_k(\bs, \bs') = \mu_k(\bs')-\mu_k(\bs) = \frac{f_k(\bs')-f_k(\bs)}{\wk}\left( e^{i\wk t} - 1 \right),
\end{equation}
are $c$-numbers and 
\begin{equation}\label{W}
W_t(\bs) = \sum\limits_{k=1}^K \frac{f_k(\bs)^2}{\omega_k^2} (\sin\omega_k t - \omega_k t).
\end{equation}
Substituting this result in Eq.~(\ref{Lambda}) then yields
\begin{equation}\label{Lambda2}
    \Lambda_t(\bs,\bs') = i \left[W_t(\bs') - W_t(\bs) \right] - \Gamma_t(\bs,\bs'),
\end{equation}
with
\begin{equation}\label{Gamma}
    \Gamma_t(\bs,\bs') = - \log{\left\langle \prod_k D_k\left(\mu_k(\bs,\bs')\right)\right\rangle_B}.
\end{equation}
Eq.~(\ref{Lambda2}) provides the general  time-evolution for each matrix element of $\rho_S$ due to the open-system map [Eq.~(\ref{local_map_elements})]. 

The first term in Eq.~(\ref{Lambda2}) is actually a unitary contribution.
This can be seen more clearly by returning to Eq.~(\ref{local_map_elements}) and writing $W_t(\bs)|\bs\rangle = W_t(\bs_z) |\bs\rangle$, which leads to 
\begin{equation}\label{local_map_elements_Lamb}
\langle \bs | \mathcal{E}(\rho_S) | \bs' \rangle = \langle \bs | e^{-i W_t(\bs_z)} \rho_S  e^{i W_t(\bs_z)} | \bs' \rangle e^{-\Gamma_{t}(\bm{\sigma}, \bm{\sigma}')}.
\end{equation}
The operator $W_t(\bs_z)$ therefore enters as a time-dependent Hamiltonian evolution. 
It therefore corresponds to a type of dynamical Lamb-shift. 
As we will show, this term can actually lead to quite rich unitary couplings because, as seen in Eq.~(\ref{W}), in depends on $f_k(\bs)^2$.
Thus, a linear interaction such as~(\ref{f_linear}) would given a $W_t(\bs_z)$ containing Ising like couplings $\sigma_z^i\sigma_z^j$. 
The ramifications of this will be explored in more detail in Sec.~\ref{sec:cases}. 
We also emphasize that this Lamb-shift is entirely independent of the initial state of the bosons, but depends only on the spin-boson coupling. 

\subsection{Structure of $\Gamma_t(\bs,\bs')$}

We now turn to the function $\Gamma_t(\bs,\bs')$ in Eq.~(\ref{Gamma}).
Notice  how it cannot be split in two terms like the unitary term in~(\ref{Lambda}). 
It therefore corresponds to a genuinely dissipative contribution. 
This quantity turns out to be related to the characteristic function of the initial bosonic state. 
Let $\bm{r} = (q_1,p_1, \ldots, q_K, p_K)$, where $q_i = (\alpha_i + \alpha_i^*)/\sqrt{2}$ and $p_i = i (\alpha_i^* - \alpha_i)/\sqrt{2}$. 
The characteristic function of $\rho_B$ is defined as~\cite{Serafini2017} 
\begin{equation}\label{chi_def}
\chi_B(\bm{r}) = \left\langle \prod\limits_k D_k(\alpha_k) \right\rangle_B. 
\end{equation}
Comparing with Eq.~(\ref{Gamma}) we therefore conclude that 
\begin{equation}\label{Gamma_chi}
\Gamma_t(\bs,\bs') = -  \ln \chi_B(\bm{r}), 
\end{equation}
evaluated at the specific phase-space points $\alpha_k = \mu_k(\bs,\bs')$ [Eq.~(\ref{mu_k})]. 
Or, more explicitly in terms of the entries of $\bm{r}$,
\begin{IEEEeqnarray}{rCl}
    q_k &=& \frac{\mu_k(\bs,\bs') + \mu_k(\bs,\bs')^*}{\sqrt{2}} \nonumber \\
    &=& \frac{\sqrt{2}}{\wk} [f_k(\bs') - f_k(\bs)](\cos{\wk t} - 1), 
    \label{qk_rep}	\\[0.2cm]
    p_k &=& \frac{i}{\sqrt{2}} (\mu_k(\bs,\bs')^* - \mu_k(\bs,\bs')) \nonumber \\
    &=& \frac{\sqrt{2}}{\wk} [f_k(\bs') - f_k(\bs)] \sin{\wk t}.
    \label{pk_rep}
\end{IEEEeqnarray}

This connection between $\Gamma_t(\bs,\bs')$ and the characteristic function $\chi_B(\bm{r})$ is general and holds for an arbitrary initial state $\rho_B$.
In order to explore which typical  structures  may emerge for $\Gamma_t$, however, it is convenient to make additional assumptions about $\rho_B$. 
First, let us suppose that $\rho_B$ is a generic Gaussian state. 
In this case the characteristic function can be expressed solely in terms of the first and second moments of  $\rho_B$.  
Define the quadrature operators $\hat{q_k} = (b_k + b_k^\dagger)/\sqrt{2}$ and $\hat{p_k} = i(b_k^\dagger - b_k)/\sqrt{2}$, as well as the vector  $\hat{\boldsymbol{r}} \equiv (\hat{q_1}, \hat{p_1}, \hat{q_2}, \hat{p_2}, \cdots, \hat{q_K}, \hat{p_K})$.
The first moments are then $\langle \hat{q}_k\rangle$ and $\langle \hat{p}_k\rangle$, whereas the second moments can be condensed onto the $2K\times 2K$ Covariance Matrix (CM):
\begin{equation}\label{CM}
\Theta_{k,k'} = \frac{1}{2}\langle \{ \hat{r}_k, \hat{r}_{k'} \} \rangle - \langle \hat{r}_k \rangle\langle \hat{r}_{k'} \rangle.
\end{equation}
The characteristic function for a Gaussian state is then~\cite{Serafini2017} 
\begin{equation}\label{chi_gaussian}
 \ln \chi_B(\bm{r}) 		= - \frac{1}{2} \boldsymbol{r}^T \Theta \boldsymbol{r} + i \bm{r}\trans \Omega \bar{\bm{r}}, 
\end{equation}
where $\bar{\bm{r}} = (\langle \hat{q}_1\rangle, \langle \hat{p}_1 \rangle, \ldots)$ is the vector of first moments and 
\begin{equation}
\Omega = \begin{pmatrix}  0 & 1 \\ -1 & 0 \end{pmatrix}^{\oplus K},
\end{equation}
is the symplectic form.  
According to Eq.~(\ref{Gamma_chi}), the function $\Gamma_t(\bs,\bs')$ will be given by Eq.~(\ref{chi_gaussian}) evaluated at the specific phase-space points~(\ref{qk_rep}) and (\ref{pk_rep}).

The last term in Eq.~(\ref{chi_gaussian}) turns out to also lead to a unitary contribution. 
In fact, using Eqs.~(\ref{qk_rep}) and (\ref{pk_rep}) one may verify that 
\begin{equation}
i \bm{r}\trans \Omega \bar{\bm{r}} = i \bigg[ \tilde{W}_t(\bs') - \tilde{W}_t(\bs)\bigg], 
\end{equation}
where 
\begin{equation}\label{W_tilde}
\tilde{W}_t(\bs) = \sqrt{2} \sum\limits_k \frac{f_k(\bs)}{\omega_k} \bigg[ \langle \hat{p}_k \rangle (\cos\omega_k t - 1) + \langle \hat{q}_k\rangle \sin\omega_k t\bigg]. 
\end{equation}
This therefore corresponds to an additional dynamical Lamb-shift, which is related to initial displacements of the bosons, $\langle \hat{p}_k\rangle$ and $\langle \hat{q}_k\rangle$. 
Contrary to the term $W_t$ in Eq.~(\ref{W}), however, this contribution yields terms which are linear in $f_k(\bs)$. 
Moreover, it depends on the initial conditions of the bosons, unlike~(\ref{W}), which exists for any initial state $\rho_B$. 
 For clarity of presentation, however, we shall henceforth omit this additional Lamb-shift by assuming that $\langle \hat{q}_k\rangle = \langle \hat{p}_k\rangle = 0$. 

Turning then to the first term in Eq.~(\ref{chi_gaussian}), it is interesting to note how it  contemplates the possibility to have initial correlations among the bosonic modes,  encoded in the covariance matrix~(\ref{CM}). 
This could be used, for instance, to study non-Markovianity and information flows in the dynamics of the qubits. 
As an additional simplification, however, we shall assume that the bosonic modes start in a product state, so that the CM is block-diagonal, with each mode having its own $2\times 2$ covariance matrix $\Theta_k$.
This simplifies Eq.~(\ref{chi_gaussian}) to $\ln \chi_B(\bm{r}) = - (1/2) \sum_k \boldsymbol{r}^T_k \Theta_k \boldsymbol{r}_k$. 
A general form for the CM $\Theta_k$ of each mode is a thermal squeezed state of the form  
\begin{equation}
  \Theta_k = (\bar{n}_k+1/2) \left(
\begin{array}{cc}
    e^{2z}/2 & 0 \\
    0 & e^{-2z}/2
\end{array}
\right), 
\end{equation}
where $\bar{n} >0$ is the Bose-Einstein occupation and $z > 0$ is the squeezing parameter, which is taken to be real for simplicity. 
Using this, as well as the specific forms~(\ref{qk_rep}) and (\ref{pk_rep}), we then arrive at 
\begin{IEEEeqnarray}{rCl}
\label{Gamma_squeezed}
    \Gamma_t(\bs,\bs') = \sum_k \frac{1}{4\wk ^2} [f_k(\bs') - f_k(\bs)]^2 \coth{\frac{\wk}{2T}} \nonumber \\ [0.2cm]
   \times \left[ (\cos{\wk t} - 1)^2 e^{2z} + \sin^2\wk t \; e^{-2z} \right] .
\end{IEEEeqnarray}
This is explicit form of the decoherence rate in the case where the modes start uncorrelated, in thermal squeezed states. 
For a single qubit, $f_k(\bs) = \lambda \sigma_1$, this reduces to the results derived recently in Ref.~\cite{You2017}. 
Eq.~(\ref{Gamma_squeezed}) therefore provides a substantial generalization, as it contemplates an arbitrary number of qubits with arbitrary interactions.
When the modes are purely thermal ($z =0$) it simplifies further to 
\begin{IEEEeqnarray}{rCl}
\label{Gamma_thermal}
    \Gamma_t(\bs,\bs') = \sum_k \frac{1}{2\wk ^2} [f_k(\bs') - f_k(\bs)]^2 \coth{\frac{\wk}{2T}} \nonumber \\ [0.2cm]
   \times \left[ 1- \cos{\wk t} \right] ,
\end{IEEEeqnarray}
which is the generalization of Eq.~(\ref{SB_Gamma}).

To summarize, the contact with the bosonic modes will cause the spins to evolve according to the map~(\ref{local_map_elements}). 
This map contains a unitary Lamb-shift contribution $W_t(\bs)$, given by Eq.~(\ref{W}), as well as a global dephasing $\Gamma_t(\bs,\bs')$ which can be written quite generally in terms of the characteristic function of $\rho_B$ according to Eq.~(\ref{Gamma_chi}). 
This formula simplifies for several particular cases, including squeezed thermal bosons [Eq.~(\ref{Gamma_squeezed})] or simply thermal bosons [Eq.~(\ref{Gamma_thermal})]. 
Possible initial displacements of the bosonic modes can also lead to an additional Lamb-shift contribution in Eq.~(\ref{W_tilde}), which we henceforth assume to be zero for simplicity.

\subsection{Evolution of the bosonic modes}

Before analyzing particular cases and examples, we briefly discuss how the evolution takes place from the perspective of the bosons. 
In particular, returning to the global map~(\ref{global_map}) and tracing instead over the qubit system, we find that the evolution of the bosonic modes is given by 
\begin{equation}
\rho_B(t) = \sum\limits_{\bs} \langle \bs | \rho_S | \bs \rangle e^{- i V(\bs) t} \rho_B e^{ i V(\bs) t}. 
\end{equation}
The evolution of the bosons therefore depend only on the initial populations of $\rho_S$ and not on the coherences. 
Thus, while the bosons affect the coherences of $\rho_S(t)$ and not its populations, $\rho_B(t)$ itself is only affected by the populations. 

To provide an example of this effect, if the bosons all start in the vacuum, we find 
\begin{equation}
\rho_B(t) = \sum\limits_{\bs} \langle \bs |\rho_S | \bs \rangle \prod\limits_k |\mu_k(\bs) \rangle\langle \mu_k(\bs)|,
\end{equation}
where $|\mu_k(\bs)\rangle$ are coherent states with the values given in Eq.~(\ref{mu_k}). 
To make this result even more intuitive, consider the case discussed in Eq.~(\ref{Dicke}) of a  single boson ($K = 1$) interacting linearly with $N$ qubits. 
In this case the state of the boson will be given by 
\begin{equation}
\rho_B(t) = \sum\limits_{\bs} \langle \bs |\rho_S | \bs \rangle  |\mu(\bs) \rangle\langle \mu(\bs)|,
\quad
\mu(\bs) = \frac{\lambda \mathcal{M}(\bs)}{\omega} (e^{i\omega t} - 1).
\end{equation}
Here we also defined the total magnetization 
\begin{equation}\label{magnetization}
\mathcal{M}(\bs) = \sum\limits_{i=1}^N \sigma_i . 
\end{equation}
The single boson will therefore evolve as an incoherent mixture of coherent states, each oscillating with the same frequency but having different magnitudes proportional to $\mathcal{M}(\bs)$.

%
%
\section{\label{sec:cases}Particular case: linear interactions}
%
%

In this section we specialize the results of Sec.~\ref{sec:sol} to the case where $f_k(\bs)$ represents a linear interaction of the form~(\ref{f_linear}).
The quantities we wish to study are the two terms in $\Lambda_t(\bs,\bs')$ [Eq.~(\ref{Lambda2})]. 
First the dynamical Lamb-shift $W_t(\bs_z)$ in Eq.~(\ref{W}), becomes 
\begin{equation}\label{W_linear}
W_t(\bs_z) = \sum_{i,j} W_{ij}(t) \; \sigma_z^i \sigma_z^j,
\end{equation}
where
\begin{equation}\label{W_ij}
W_{ij}(t) =   \sum_{k=1}^K \frac{\lambda_{ki} \lambda_{kj} }{\omega_k^2} (\sin \omega_k t - \omega_k t). 
\end{equation}
The Lamb-shift therefore acquires the form of a time-dependent classical Ising interaction. 
The coupling between qubits $i$ and $j$ is determined by  the product $\lambda_{ki} \lambda_{kj}$; 
it therefore depends on which bosons $k$ are coupled to both qubits so as to mediate an effective interaction. 
By choosing which qubits interacts with which bosons (Fig.~\ref{fig:diagram}), one can therefore construct an arbitrary network of Ising couplings among the qubits.


Next we turn to the dephasing rate~(\ref{Gamma}). 
For concreteness, we shall focus on the case of  thermal bosons, Eq.~(\ref{Gamma_thermal}). 
Substituting~(\ref{f_linear}) for $f_k(\bs)$  yields 
\begin{equation}\label{Gamma_linear}
\Gamma_t(\bs,\bs') = \sum_{i,j} \Gamma_{ij}(t)(\sigma_i'-\sigma_i)(\sigma_j'-\sigma_j),
\end{equation}
where 
\begin{equation}\label{Gamma_ij}
\Gamma_{ij}(t) = \sum\limits_{k=1}^K \frac{\lambda_{ki} \lambda_{kj} }{2\omega_k^2} \coth\left(\frac{\omega_k}{2T}\right) (1- \cos\omega_k t).
\end{equation}
One notices that the time-dependence of $W_{ij}$ and $\Gamma_{ij}$ are fundamentally different. 
The former has a term which is linear in $\omega_k t$, whereas the latter only contains an oscillatory contribution. 
Changing the values of $\lambda_{ki}$ and $\omega_k$ one we can thus find time intervals where the dephasing term $\Gamma_{ij}$ can become negligible compared with to Lamb-shift $W_{ij}$, or vice-versa. 
Thus, for certain intervals, the dynamics can be approximately almost unitary.

\subsection{Local environments}

The structure of Eq.~(\ref{Gamma_linear}) contains a quite rich physical behavior. 
In order to better appreciate it, we consider two limiting cases. 
First, suppose  each qubit only interacts with its own set of bosonic modes. 
As discussed in Sec.~\ref{sec:int}, this can be introduced by sectorizing the interactions, so that $\lambda_{k1} \neq 0$ only for $k = 1, \ldots, m$ (where $m$ is an integer), $\lambda_{k2} \neq 0$ only for $k = m+1, \ldots, 2m$ and so on. 
This implies that in both~(\ref{W_ij}) and (\ref{Gamma_ij}) there will be no terms for which $\lambda_{ki} \lambda_{kj} \neq 0$ when $j \neq i$. 
Consequently, $W_{ij}$ and $\Gamma_{ij}$ will  both be diagonal. 
The unitary contribution~(\ref{W_linear}) in this case is just a constant (since $(\sigma_z^i)^2 = 1$) and therefore does not contribute at all to the dynamics. 
The dephasing term~(\ref{Gamma_linear}), on the other hand, reduces to 
\begin{equation}\label{Gamma_linear_local}
\Gamma_t(\bs,\bs') = \sum\limits_i \Gamma_{ii}(t) (\sigma_i' - \sigma_i)^2.
\end{equation}
Let us also suppose, for the sake of argument, that all spins are coupled to an identical set of bosons, so that $\Gamma_{ii}$ in Eq.~(\ref{Gamma_ij}) becomes independent of $i$:
\begin{equation}\label{Gamma_ij_local}
\Gamma(t) = \sum\limits_{k=1}^m \frac{\lambda_k^2}{2\omega_k^2} \coth\left(\frac{\omega_k}{2T}\right) (1- \cos\omega_k t).
\end{equation}
In this case we can write~(\ref{Gamma_linear_local}) in terms of the magnetization operator~(\ref{magnetization}) as 
\begin{equation}\label{Gamma_linear_local_uniform}
\Gamma_t(\bs,\bs') = \Gamma(t) \bigg( \mathcal{M}(\bs') - \mathcal{M}(\bs) \bigg)^2. 
\end{equation}
This result can be connected with the notion of coherence orders~\cite{Pires2017}. 

The coherences in the density matrix $\rho_S$ can be divided into sectors, called coherence orders, corresponding to different total magnetizations. 
To give a concrete example, consider $N=6$ qubits prepared in the superposition  state 
\[ 
|\psi_S\rangle = \frac{|1,1,1,1,1,1\rangle + |1,1,1,-1,1,1\rangle}{\sqrt{2}}.
\]
The coherences present in $\rho_S = |\psi_S\rangle\langle \psi_S|$ will, in this case, be between a sector with $\mathcal{M} = 6$ and $\mathcal{M} = 4$. 
Conversely, consider a superposition of the form 
\[ 
|\psi_S\rangle = \frac{|1,1,1,1,1,1\rangle + |-1,-1,-1,-1,-1,-1\rangle}{\sqrt{2}}.
\]
This superposition is much more dramatic. 
It contains coherences between  macroscopically distinct states, with $\mathcal{M} = 6$ and $\mathcal{M} = -6$.
In the limit of macroscopically large $N$, for instance, this would be tantamount to preparing a magnet in a superposition of all spins up and all spins down. 

Recall that $\Gamma_t(\bs,\bs')$ appears multiplying the corresponding coherence as $e^{-\Gamma_t(\bs,\bs')}$ [Eq.~(\ref{local_map_elements_Lamb})]. 
The result in Eq.~(\ref{Gamma_linear_local_uniform}) therefore shows that this decoherence rate will be proportional to the net magnetization difference between $|\bs\rangle$ and $|\bs'\rangle$. 
For configurations   with drastically distinct values of $\mathcal{M}$, this will therefore lead to exponentially fast dephasing rates. 
This therefore provides a beautiful (and exact) example of einselection~\cite{Zurek1981,Zurek2003b}: macroscopic quantum features are much more susceptible to the effects of the environment and hence degrade extremely fast. 

In the case where the spin couplings are not uniform, the dephasing rate~(\ref{Gamma_linear_local}) cannot be written as elegantly as~(\ref{Gamma_linear_local_uniform}), in terms solely of $\mathcal{M}(\bs)$. 
Notwithstanding, the logic remains the same: states with dramatically different spin configurations tend to have exponentially larger decoherence rates.

\bigskip
\subsection{Fully connected model}

As another limiting case, we consider the situation where all spins are coupled to all bosons in exactly the same way. 
This means that $\lambda_{ki} = \lambda_k$ are independent of $i$ for all $k \in [1,K]$. 
As a consequence, Eqs.~(\ref{W_linear}) and (\ref{W_ij}) become
\begin{equation}\label{W_linear_fully}
W_t(\bs_z) = W(t) \mathcal{M}(\bs_z)^2, 
\end{equation}
where
\begin{equation}\label{W_ij_fully}
W(t) = \sum\limits_k \frac{\lambda_k^2}{\omega_k^2} (\sin\omega_k t - \omega_k t). 
\end{equation}
The Lamb-shift therefore yields an interaction proportional to the net magnetization operator $\mathcal{M}(\bs_z)$, akin to the typical interaction appearing in the Lipkin-Meshkov-Glick model~\cite{Ribeiro2008}.

The dephasing rate~(\ref{Gamma_linear}), on the other hand, becomes exactly like Eq.~(\ref{Gamma_linear_local_uniform}), but with 
\begin{equation}\label{Gamma_ij_fully}
\Gamma(t) = \sum\limits_{k=1}^K \frac{\lambda_k^2}{2\omega_k^2} \coth\left(\frac{\omega_k}{2T}\right) (1- \cos\omega_k t).
\end{equation}
The only difference with respect to Eq.~(\ref{Gamma_ij_local}) is that the sum here is over all Bosons. 
Thus, the basic structure of the dephasing rate remains unchanged in this case, when compared with local environments. 
What changes fundamentally is the appearance of the Lamb-shift contribution~(\ref{W_linear_fully}).

\subsection{Macroscopically large environments}

The results discussed so far always assume that the number of bosonic modes $K$ is arbitrary. 
In this last section we  indicate how these results change when the environment is macroscopically large so that all sums can be converted to integrals. 
For concreteness, we focus on the fully connected case, Eqs.~(\ref{W_ij_fully}) and (\ref{Gamma_ij_fully}), since these represent the typical kinds of Lamb-shifts and decoherence rates appearing in this model. 
Following standard treatments of open quantum systems, we introduce the spectral density 
\begin{equation}
J(\omega) = \sum\limits_{k=1}^K \lambda_k^2 \; \delta(\omega - \omega_k). 
\end{equation}
so that Eqs.~(\ref{W_ij_fully}) and (\ref{Gamma_ij_fully}) can be written as 
\begin{IEEEeqnarray}{rCl}
\label{W_int}
W(t) 			&=& \int\ud \omega \; \frac{J(\omega)}{\omega^2} (\sin \omega t - \omega t),		\\[0.2cm]
\Gamma(t) 	&=& \int\ud \omega \; \frac{J(\omega)}{\omega^2} \coth\left(\frac{\omega}{2T}\right) (1- \cos\omega t). 
\label{Gamma_int}
\end{IEEEeqnarray}
Thus, all results shown in the previous section are readily generalized for infinite baths. 
In order to actually carry out the  integrals, one must  of course provide a specific form for the spectral density (see~\cite{Breuer2007} for an example). 
Finally,  the case where $\lambda_{ki}$ depends on the spin $i$ can be treated in a similar way by defining a set of spectral densities 
\begin{equation}
J_{ij} (\omega) = \sum\limits_k \lambda_{ki} \lambda_{kj} \; \delta(\omega - \omega_k). 
\end{equation}

\section{\label{sec:disc} Discussion}
%
%

In this paper we have put forth a general solution for a broad class of pure-dephasing spin-boson models. 
The basic assumptions in our model are that the $N$ spins interact with the $K$ bosons only through interactions which commute with the spin Hamiltonian $H_S$. 
This implies that the bosons cannot cause any transitions among the energy states of the spins. 
Notwithstanding, they can still cause decoherence. 
In the multipartite scenario, decoherence becomes much less trivial, as our model shows. 

First, there is the appearance of non-trivial unitary Lamb-shifts which represent the effective qubit-qubit interactions mediated by the bosons. 
For the case of linear spin-boson couplings, this interaction is of the Ising type [e.g. Eq.~(\ref{W_linear})]. 
Such an interaction is frequently used, for instance, in the preparation of Hamiltonian graph states~\cite{Hein2004}.
In this case one may, for instance, prepare the qubits in a state $|\psi_S\rangle = |+\rangle^{\otimes N}$, where $|+\rangle = (|0\rangle + |1\rangle)/\sqrt{2}$. 
The Lamb-shift~(\ref{W_linear}) will, in this case, lead to a highly non-trivial, and generally entangling, evolution. 
Of course, this occurs concomitantly with decoherence. 
However, as we have shown, one can tune the spin-boson interactions so as to maximize the Lamb-shift as compared with the dephasing. 

The second non-trivial effect which appears due to the multipartite nature of our model is the emergence of  inhomogeneous decoherence rates. That is, rates which are different depending on the type of coherence involved. 
The coherences can be split into different coherence orders, associated with the difference in net magnetization of the different sectors. 
And, as shown for instance by Eq.~(\ref{Gamma_linear_local_uniform}) the decoherence rate will be larger for higher coherence orders. 
As a consequence, superpositions of macroscopically distinct states tend to be suppressed much more quickly. 

The above arguments show how our result can be used to shed light on several relevant questions about the nature of open quantum systems in the multipartite scenario, all of which accomplished within an exactly soluble model. 
In addition to this more fundamental contribution, we believe our results can also be exploited for several applications. 
One, for instance, is the just mentioned use of the Lamb-shift~(\ref{W_linear}) in the preparation of Hamiltonian graph states.
The general structure of Eqs.~(\ref{W_ij}) and (\ref{Gamma_ij}) offer general guidelines on how to use bosons to mediate Ising interactions, while at the same time minimizing potential decoherence effects.
Interactions more complicated than linear could also in principle be used, by simply changing the choice of the function$ f_k(\bs_z)$ in Eq.~(\ref{V}). 

Another potential application is in quantum metrology. 
Spin-boson models offer an interesting alternative for phase estimation \cite{Walborn2018,Razavian2018}, in which quantities such as the Fisher information can be computed exactly for arbitrary interaction times and strengths. 
The generalizations put forth in this paper allow one to extend these results to the multipartite scenario. 
One example is thermometry, where we could use the $N$ qubits to estimate the temperature of the bosons.
The analysis for a single qubit was done recently in~\cite{Razavian2018}. 
The extension to multiple qubits would allow one to exploit how collective features (e.g., from different initial states $\rho_S$) could be used to outperform individual qubit thermometry.

\section*{Acknowledgements }
The authors would like to thank Rafael Chaves, Jader P. Santos, Diogo Soares-Pinto and Diego Paiva Pires for fruitful discussions. 
The authors also acknowledge the International Institute of Physics, where part of this work was developed, for both the hospitality and the financial support. 
This work was partially funded by the University of S\~ao Paulo, the S\~ao Paulo Research Foundation FAPESP (grant numbers 2016/08721-7  and 2017/20725-0), and the Brazilian funding agency CNPq (grant number INCT-IQ 246569/2014-0).


\bibliography{/Users/gtlandi/Documents/library}
\end{document}